\documentclass[
preprint,
 amsmath,amssymb,
 aps,
]{revtex4-2}

\usepackage{graphicx}% Include figure files
\usepackage{dcolumn}% Align table columns on decimal point
\usepackage{bm}% bold math
\usepackage[utf8]{inputenc}
\usepackage[mathlines]{lineno}% Enable numbering of text and display math
\begin{document}

\preprint{}

\title{Modeling the dynamical behavior of memristive {NiTi} alloy\\ at constant stress for time-varying electric current input signals}

% Force line breaks with \\

\author{Ioannis P. Antoniades \thanks{Corresponding author. Email: iantoniades@auth.gr }}
 \email{iantoniades@auth.gr}
\affiliation{%
 Department of Physics, Aristotle University of Thessaloniki, Greece}%

\author{Stavros G. Stavrinides}
%\homepage{http://www.ihu.edu.gr/}
\affiliation{
 School of Science and Technology, 
 International Hellenic University, 
 Greece}%

\author{Rodrigo Picos}
\affiliation{%
Physics Department, 
 Universitat de les Illes Balears, Spain
 \\
 Balearic Islands Health Research Institute (IdISBa), Spain}
 
\author{Michael P. Hanias}
\affiliation{%
Physics Department, International Hellenic University, Greece}
 
\author{Mohamad Moner Al Chawa}
\affiliation{Institute of Circuits and Systems, Technische Universität, Germany
%Dresden, Dresden, 01069
}

\author{Julius Georgiou}
\affiliation{
 Department of Electrical and Computer Engineering, University of Cyprus, Cyprus}%
 
\author{Euripides Hatzikraniotis}
\affiliation{
 Physics Department, Aristotle University of Thessaloniki, Greece}%

\author{Leon O. Chua}
\affiliation{%
Electrical Engineering Department, University of California, Berkeley, CA, USA
 }%


\date{\today}% It is always \today, today,

\begin{abstract}
The dynamical electric behavior of a NiTi smart alloy thin filament when driven by time varying current pulses is studied by a structure-based phenomenological model that includes rate-based effects. The simulation model relates the alloy's electrical resistivity to the relative proportions of the three main structural phases namely Martensite, Austenite and R-phase, experimentally known to exist in NiTi alloy lattice structure. The relative proportions of the phases depend on temperature and applied stress. Temperature varies due to the self-heating of the filament by the Joule effect when a current pulse passes and also due to convective/radiative interchange with the ambient. The temperature variation with time causes structural phase transitions, which result in abrupt changes in the sample resistivity as the proportions of each lattice phase vary.% with time.
The model is described by a system of four 1st-order nonlinear differential-algebraic equations yielding the temporal evolution of resistivity and output voltage across the filament for any given time-varying input current pulse. The model corresponds to a 4th-order extended memristor, described by four state variables, which are the proportions of each of the three NiTi lattice phases and temperature.
Simulations are experimentally verified by comparing to measurements obtained for samples self-heated by triangular current input waveforms as well as for passively samples with no current input. Numerical results reproduce very well measurements of resistance vs. temperature at equilibrium as well as the full dynamics of experimentally observed I-V characteristic curves and resistance vs. driving current for time-varying current input waveforms of a wide range of frequencies (0.01-10~Hz). 

%\begin{description}
%\end{description}
\end{abstract}

 %Use showkeys class option if keyword
\maketitle

%\tableofcontents

\section{\label{sec:intro}Introduction}

Nickel-Titanium (NiTi) alloy is a member of a broad family of metals like copper-aluminum-nickel,copper-zinc-aluminum-nickel etc. called \textit{smart alloys}, each of them demonstrating diverse and noteworthy properties. Specifically, NiTi exhibits two unique properties; the \textit{shape memory effect} and \textit{super-elasticity}. NiTi-based materials are widely accepted as some of the best families of Shape-Memory Alloys (SMA). SMA materials are sensitive to temperature and/or stress, producing a large macroscopic strain, through the so-called martensitic transformation. Shape memory is the ability of NiTi wires to undergo deformation at one temperature, and then recover their original, un-deformed shape upon heating above a critical temperature. This is due to the low thermal activation energy between the two lattice phases that they demonstrate for the equatomic composition.
Martensitic transformation is a thermoelastic reversible crystallographic phase transition from a 
high-temperature phase, referred to as Austenite (A-phase), to a low temperature phase, referred to as Martensite (M-phase) and vice-versa. The A-phase is made of a simple cubic crystal structure, M-phase is a more complicated monoclinic (B19') structure.

Until now, NiTi alloys have been used mainly for their mechanical characteristics, particularly the contraction of the metal under the thermo-elastically induced martensitic transformation. This means that NiTi alloys demonstrate an extraordinary ability to recall (by returning to) a trained shape, once heated after being plastically deformed. What is noteworthy is the fact that it is possible to adjust the temperature and stress thresholds where this occurs by varying the composition of the alloy; in other words, the recall temperature and stress levels can be predetermined by the stoichiometric ratio of nickel to titanium during manufacturing.

\begin{figure}[b]
    \centering
    \includegraphics[scale=0.35]{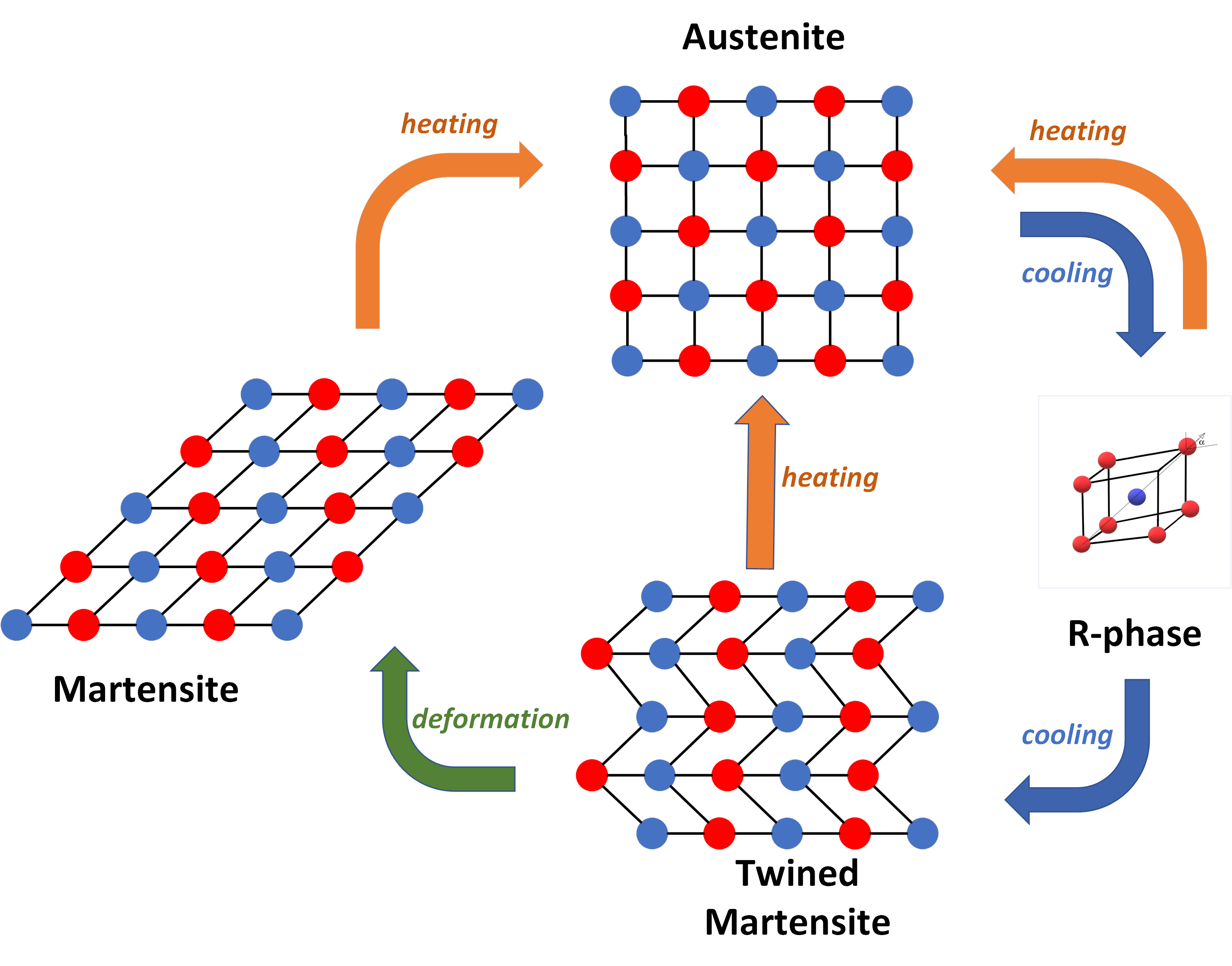}
    \caption{The three phases of the NiTi lattice structure and the transitions among them as considered in the present work.}
    \label{fig:phase_change_1}
\end{figure}

On the other hand, the electrical properties of NiTi are also important to study by projecting its structural properties to its electric behavior. It should be noted that these alloys are conductors with a high electrical resistance, which allows for substantial Joule heating. In a NiTi alloy wire, the change in the lattice structure which is induced by a change in temperature and/or applied stress, causes a change in electrical resistance. Although both M-phase and A-phase have approximately the same \textit{resistivities}, A-phase resistivity being slightly higher, a phase transformation between the low-temperature M-phase and the high-temperature A-phase will cause an abrupt change in the sample's \textit{resistance} at some critical temperature, due to an abrupt change in strain (in the order of 8-10\%). The transition may be driven by passive heating (i.e. by changing the temperature of the environment) or by self-heating caused by an electric current flowing through the sample. If the sample is then cooled down, the A-phase is first converted to a \textit{twinned} Martensite structure, as shown in fig.~\ref{fig:phase_change_1} and then, after deformation caused by an externally applied stress, it is converted back to the monoclinic (deformed) M-phase at low temperatures causing the strain to increase again to the original value and thus, the resistance to rise. The critical temperature for the reverse transformation, from A-phase to twinned Martensite, that occurs during cooling, can be different from the critical temperature of the forward transition that occurs during heating.

\begin{figure}[b]
    \centering
    \includegraphics[scale=0.4]{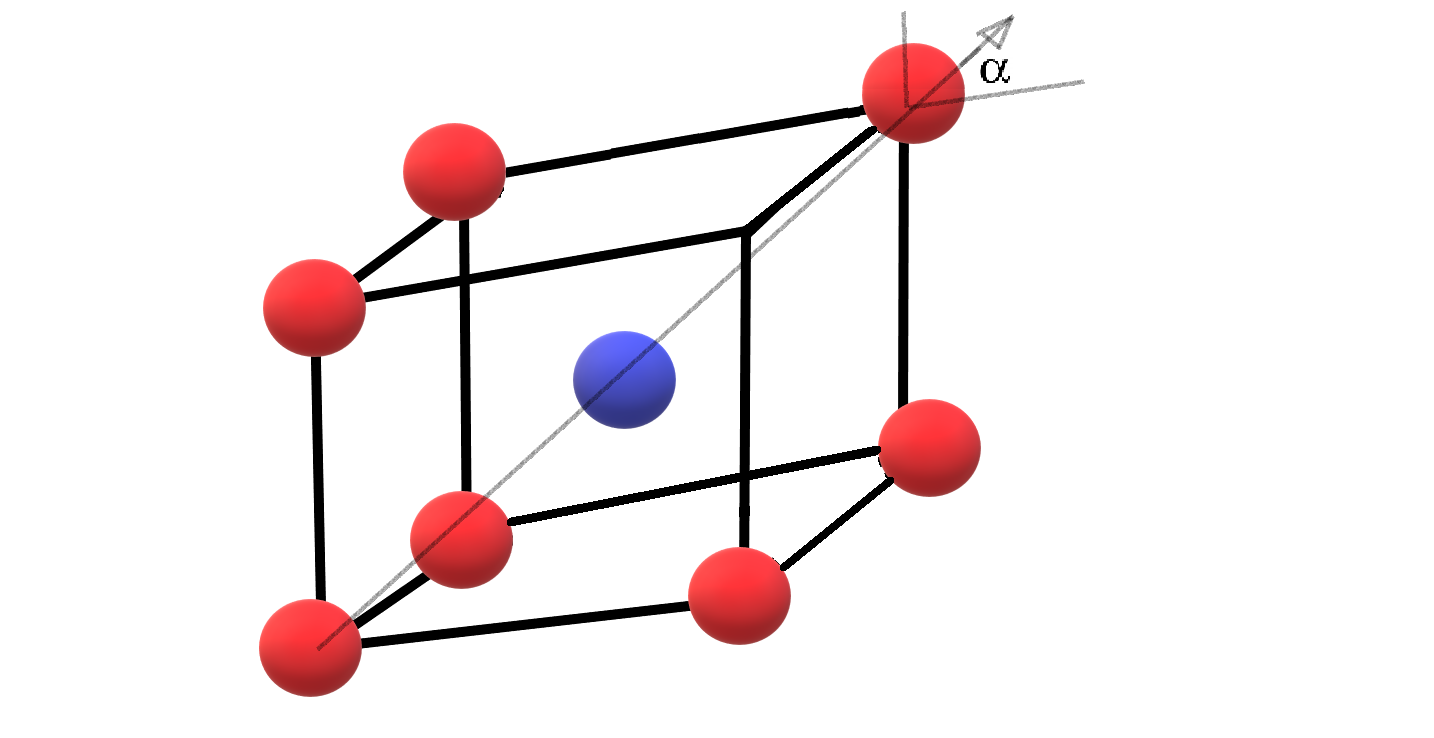}

    \caption{Structure a NiTi R-phase lattice cell. The distortion angle $\alpha$ is shown.}
    \label{fig:R-phase}
\end{figure}

Focusing on the NiTi sample \textit{resistivity} rather than resistance dependence on temperature one should first notice, as mentioned above, that M-phase (both twinned and deformed) has only slightly lower resistivity than A-phase, if projected to \textit{the same temperature value}. However, as M-phase is stable at much lower temperatures than A-phase, its resistivity is expected to be much lower, due to the linear resistivity decrease with temperature valid for any metal. Therefore, the resistivity of a NiTi sample consisting solely of M-phase, if heated above the Austenitic transition threshold, is expected to rise and similarly drop, if cooled back down to temperatures at which M-phase is stable. However, several experimental studies have shown that, NiTi resistivity rises during cooling and drops during heating at temperatures around the Austenitic transition (e.g. \cite{song2011resistance, Gori06, bhargaw2013thermo}). This anomalous behavior has been explained by the existence of an unstable intermediate NiTi lattice phase, called R-phase \cite{Ling1981, Miyazaki1986, OTSUKA1999511}. R-phase usually emerges only from Austenite during cooling. The lattice structure of R-phase is a deformed simple cubic lattice in the direction [110], as shown in fig.~\ref{fig:R-phase}, where the angular deformation $\alpha$ is rather small ($\alpha \approx 1^\circ$). It has also been reported in experimental studies that, during cooling below some transition temperature, either only R-phase is formed or it is simultaneously formed with twinned Martensite (e.g. \cite{Duerig2015}). Although the R-phase resistivity and its temperature dependence is very similar to that of A-phase, its value is rather sensitive to even slight changes in the deformation angle $\alpha$. Since further cooling of the sample causes $\alpha$ to change smoothly, the R-phase resistivity sharply rises to values significantly higher than that of Austenite. Based on the relative proportion of R-phase to the other two phases, the \textit{total} resistivity (and resistance) of the NiTi sample may then rise during cooling, instead of dropping, which would be the case if only twinned Martensite were formed, as the resistivity of twinned Martensite being slightly smaller than that of A-phase. After further cooling, below some other transition temperature, R-phase eventually converts back to Martensite, causing the sample resistivity to drop again \cite{Gori06}. During heating, R-phase also converts to A-phase, after an equivalent reverse change of $\alpha$. Although R-phase converts to M-phase during cooling, M-phase usually does not convert to R-phase during heating. For a thorough discussion of how transitions may occur from one phase to the other, the reader is diverted to the background section of the paper by Duerig and Bhattacharya \cite{Duerig2015}. Ni/Ti stoichiometry, possible externally applied stress, impurities and sample treatment strongly affect the precise temperature dependence of resistivity, as these factors affect transition temperatures among phases and the relative proportion of each phase. It has also been reported that resistivity may also vary based on the \textit{order} of the heating-cooling cycle, indicating that there is some change of the lattice structure after each cycle \cite{Gori06}.

Because of the possible difference in transition temperatures between heating or cooling, the existence of \textit{hysteretic} behavior, (i.e. the difference of resistivity \textit{vs.} temperature curves between the cooling and heating parts of the cycle) has been experimentally shown. Thus, NiTi behaves effectively as an \textit{extended memristor} \cite{stavrinides2019niti, georgiou2012niti, kyriakides2012new} with the temperature change playing the role of one internal state variable and the fraction of each lattice phase acting as additional state variables. Memristors are a very interesting and promising contribution to electronic technology of the 21st century coming from the 70s considered to implement the theoretically predicted fourth basic electrical element  \cite{chua1971memristor}. This element was experimentally demonstrated much later by Strukov et al. \cite{strukov2008missing}, in 2008. However, with the insight that comes from experience, it has been proposed that memristor was actually the second basic element ever described \cite{prodromakis2013two}. The importance of the existence of properly operating memristive devices is apparent, since it would lead to a change in information and communications technology, most probably towards bio-inspired (neuro-morphic) computing.
Memristive devices can be implemented in a wide range of technologies, from spintronics \cite{grollier2016spintronic} to organic materials \cite{sun2017organic,battistoni2017organic} and many different oxides \cite{strukov2008missing, dias2017bipolar, brivio2014formation, mohammad2016state}, to mention only a few.

Regarding previous work in theoretical modeling and simulation of NiTi resistivity/resistance dependence on temperature and stress, it should be mentioned that there have been a few attempts in the direction of phenomenological models for NiTi samples at thermodynamic equilibrium (eg. \cite{NOVAK2008127,song2011resistance,Cui_2010} which are structure-based, i.e. they express the total resistivity/resistance of the sample as a linear combination of the resistivities of each lattice phase according to the volume fraction of each phase. To the best of our knowledge, these models produce the resistance/resistivity dependence on temperature and/or stress at constant temperature and they do not report rate-based effects on resistance such as those that are expected when applying time-varying current pulses

In this paper, we study the temporal evolution of electrical properties of NiTi alloy based on \textit{thermally} produced dynamical changes in NiTi lattice caused by passive heating and self-heating under time-varying currents by employing a dynamic structure-based semi-phenomenological model. As our goal is to reproduce thermally induced changes in the sample's resistivity we do not consider stress/strain related terms. The model represents the alloy as a mixture of three lattice phases (M-phase, A-phase and R-phase) and describes their temporal evolution through a system of ordinary differential equations with respect to time. Temperature is assumed to be a dynamic variable, together with the each lattice phase proportion. The temperature induced lattice transformations are described phenomenologically by using empirical functions to model the relative change in lattice phase proportions as temperature is varied. Temperature evolves in time via a heat-balance ODE. The model outputs the temporal evolution of (i) the proportions of each lattice phase, (ii) the electric resistivity of the sample, (iii) the voltage across the sample and (iv) the sample temperature itself.  Temperature is also assumed to be spatially uniform in the entire volume of the NiTi sample, as we consider only very thin NiTi wires. The system of ODE's is numerically integrated and numerical results are then compared to available experimental data for NiTi samples self-heated by periodic current pulses of various frequencies, covering a wide range (0.01 to 10~Hz). Finally, we discuss the connection of the present NiTi electric behavior dynamic simulation model to a 4th order extended memristor and the possibility to extend the model to include external (possibly time-varying) applied stress.  

\section{\label{sec:model} Model Description}

As already mentioned, the model assumes three possible lattice configurations for the atomic structure of NiTi, as shown in Fig. \ref{fig:phase_change_1}. These different phases before, after or during the temperature-induced lattice phase transformations are namely \textit{Martensite} (M), \textit{Austenite} (A), and \textit{R-phase} (R). The three lattice phases may coexist in a NiTi sample for any given temperature.
The fact that the resistivity of the R-phase may vary on deformation angle $\alpha$ is effectively treated by considering only the peak value of the resistivity of R-phase (at the maximum possible deformation), thus avoiding to treat $\alpha$ as an extra dynamic variable, as the precise temperature dependence of the variation of $\alpha$ is not precisely known. Thus, for this version of the model, we effectively define the R-phase as a lattice structure that can only exist at maximum angle deformation. The resistivity of R-phase at different deformations can be effectively described as a mixture of the maximum deformation R-phase resistivity and the resistivity of M-phase and A-phase.

Each lattice phase is assumed to have different resistivity with different dependence on temperature. Let $\rho_{M}(T), \rho_{A}(T)$ and $\rho_{R}(T)$ be the resistivity of M, A and R lattice phases in respect, as a function of (absolute) temperature $T$. As expected for any metal, the individual temperature dependence of each phase is described by a linear relationship:

\begin{eqnarray}
   \rho_{M} \left( T \right) = \rho _{M,0} \left[ 1+a_{M} \left(T-T_{0} \right)  \right] \nonumber \\
  \rho _{A} \left( T \right) = \rho _{A,0} \left[ 1+a_{A} \left( T-T_{0} \right)  \right] \nonumber \\
   \rho _{R} \left( T \right) = \rho _{R,0} \left[ 1+a_{R} \left( T-T_{0} \right)  \right]
   \label{eq:ResVsT}
\end{eqnarray}
\noindent $\rho_{M,0}(T), \rho_{A,0}(T), \rho_{R,0}(T)$ being the resistivities at a reference temperature $T_0$  and  $a_{M}, a_{A},a_{R}$ the respective thermal rates of change.
%, as shown in Fig. \ref{fig:phase_change_2}. 
Regarding phase transitions, the start and finish temperatures for a transition from lattice type A to lattice type B, in general, differs between heating and cooling. Based on available experimental results mentioned in the introduction, the allowed phase transitions are assumed to be the following: 1) Heating process: Both M-phase and R-phase lattice types convert to A-phase ($M \rightarrow A, R \rightarrow A$. 2) Cooling process:  A-phase converts to R-phase and R-phase to M-phase: $A \rightarrow R, R \rightarrow M$. The latter transitions may occur concurrently. We note that, in this paper, when we refer to M-phase we mean the twinned (un-deformed) Martensite lattice structure. the present model does not consider the transformation of twinned Martensite to deformed Martensite caused by external applied stress. In general, M-phase is known to exist at least three distinct variants, the present paper considers that in terms of resistivity these can be expressed as being one.   

As in previous studies (\cite{NOVAK2008127,Cui_2010}, the assumption that the present model makes is that, at any point in time $t$ during the transition process and at a particular temperature $T$, the effective resistivity  $\rho(T;t)$ of the NiTi sample, whose atomic structure is made up of a possible mixture of all three lattice types, is a linear combination of the respective resistivities (eq. \eqref{eq:ResVsT}) of each lattice type. Thus the total sample resistivity is given by:

\begin{equation}
 \rho  \left( T;t \right) = \sum_{i=1}^3 \xi_{i}\left( T;t \right)  \rho_{i} \left( T \right),  i \in \{M,A,R\} 
 \label{eq:SampleResistivity}
\end{equation}

\noindent where the coefficients $\xi_{i}(T;t)$ represent the fraction of each lattice type $i$ in the lattice configuration at time $t$ and temperature $T$. The above assumption ignores possible effects of texture, inter-phase and inter-granular boundaries on the sample resistance, as electron wave scattering increases around these extended defects. For a discussion on how these may influence the sample's electrical behavior see section \ref{sec:Discussion}. By definition, the $\xi$’s must satisfy the following restrictions:

\begin{eqnarray}
  \xi _{i} \left( T;t \right)  \leq 1 \nonumber \\
\sum_{i=1}^3 \xi_{i} \left( T;t \right) \equiv 1  
 \label{eq:KsiRestrictions}
\end{eqnarray}

%The variation of $\xi_{M}(T)$, $\xi_{A}(T)$ and $\xi_{R}(T)$ with temperature is determined as follows: 
The change $d\xi_{i}$ of any fraction $\xi_{i}$ (\textit{i}=M, A or R) of a parent phase $i$ after a  change in temperature $dT$ is assumed to be proportional to the fraction $\xi_{j}$ (\textit{j}=M, A or R) of the target lattice type $j$ multiplied by a function of temperature that depends on the critical start and finish temperatures, $T_{s,ij}$, $T_{f,ij}$ of the corresponding phase transition. Taking into account the allowed phase transitions and the second restriction in equation \eqref{eq:KsiRestrictions}, we have:

\begin{enumerate}
\item Heating process:
\begin{subequations}
\begin{equation}
  \frac{d \xi _{M}}{dT} = -\xi _{M}F_{M \rightarrow A} \left( T \right)
  \label{subeq:RawHeating1}
  \end{equation}
\begin{equation}
 \frac{d \xi _{A}}{dT} = \xi _{M}F_{M \rightarrow A} \left( T \right) + \xi _{R}F_{R \rightarrow A} \left( T \right)
  \label{subeq:RawHeating2}
  \end{equation}
\label{eq:RawHeating}
\end{subequations}

\item Cooling process:
\begin{subequations}
\begin{equation}
  \frac{d \xi _{M}}{dT}= \xi _{R}F_{R \rightarrow M} \left( T \right)
\label{subeq:RawCooling1}
\end{equation}
\begin{equation}
\frac{d \xi _{A}}{dT}= -\xi _{A}F_{A \rightarrow R} \left( T \right)
\label{subeq:RawCooling2}
\end{equation}
\label{eq:RawCooling}
\end{subequations}
\item Both processes:
\begin{equation}
 \xi_{R}= 1 - \xi_{M} - \xi_{A}.
 \label{subeq:XiR}
\end{equation}
\end{enumerate}

\noindent $F_{i \rightarrow j} \left( T \right)$ is the function of temperature specific for the transition from lattice type $i$ to $j$($i, j \in \{M, A, R\}$).

In order to determine the unknown functions $F_{i \rightarrow j} \left( T \right)$, \textit{limiting} cases are considered corresponding to the situation where, at time $t=0$, (i) the NiTi sample structure comprises of a single phase M, A, or R, (ii) temperature is far away from the critical start temperature of a transition from phase $i$ to $j$, (iii) temperature \textit{monotonically} increases (or decreases) at a constant rate until the initial lattice phase is \textit{fully} converted to the target lattice phase. (iv) The temperature changes slowly enough so that the system is in thermodynamic equilibrium at each temperature. There are, in total, four such limiting cases that we consider in the present model, one for each allowed transition. For determining the shape of the transition function with temperature, we follow a phenomenological approach; as one clearly sees, experimental resistivity data can accurately be fitted by a \textit{smooth sigmoid} function centered somewhere between the start and finish temperatures of transition. In the present model, the standard logistic function $S(x)=\frac{1}{1+\exp({-x})}$ is initially selected as the simplest choice, which is strictly symmetrical around its center point. As we will show next, one can update the sigmoid to a more versatile form by applying a very simple modification. For now, with $S(x)$ as the sigmoid, the dependence of the fraction of a parent phase $\xi _{i} \left( T;t \right)$ on temperature for each limiting case is described by the following equations:
\begin{itemize}
\item limiting case 1 ($M \rightarrow A$  transition, \textit{Heating}):  
 $T(t=0) \ll T_{c,MA}$, $\xi_{M}(T;t=0)=1$, $\xi_{A}(T;0)=0$, $\xi_{R}(T;0)=0$: 
 
\begin{eqnarray}
\xi _{M} \left( T;t \right) =S \left( -m_{MA} \left( T-T_{c,MA} \right)  \right) \nonumber \\
\xi _{A} \left( T;t \right) =1- \xi _{M} \left( T \right) =S \left( m_{MA} \left( T-T_{c,MA} \right)  \right) \\
\xi _{R} \left( T;t \right)  \equiv 0 \nonumber
\label{eq:LimCase1}
\end{eqnarray}

\item limiting case 2 ($A \rightarrow R$ transition, \textit{Cooling}): 
$T(t=0) \gg T_{c,AR}$, $\xi _{M}(T;0)=0$, $\xi_{A}(T;0)=1$, $\xi_{R}(T;0)=0$
\begin{eqnarray}
\xi _{M}(T;t)  \equiv 0 \nonumber \\
\xi_{A}(T;t)=S \left( -m_{AR} \left( T-T_{c,AR} \right)  \right) \\
\xi_{R}(T;t) = 1- \xi _{A} \left( T;t \right) =S \left( m_{AR} \left( T-T_{c,AR} \right)  \right) \nonumber 
\label{eq:LimCase2}
\end{eqnarray}

\item limiting case 3 ($R \rightarrow M$ transition,\textit{Cooling}):

$T(t=0)  \gg T_{c,RM}$, $\xi_{M}(T;0)=0$, $\xi_{A}(T;0)=0$, $\xi _{R}(T;0)=1$
\begin{eqnarray}
\xi _{M} \left( T;t \right) =S \left( m_{RM} \left( T-T_{c,RM} \right)  \right) \nonumber \\
\xi _{A} \left( T;t \right)  \equiv 0 \\
\xi _{R} \left( T;t \right) =1- \xi _{M} \left( T;t \right) =S \left( -m_{RM} \left( T-T_{c,RM} \right)  \right) \nonumber
\label{eq:LimCase3}
\end{eqnarray}

\item limiting case 4 ($R \rightarrow A$ transition,\textit{Heating}):

$T(t=0)  \ll T_{c,RA}$, $\xi_{M}(T;0)=0$, $\xi_{A}(T;0)=0$, $\xi _{R}(T;0)=1$
\begin{eqnarray}
\xi _{M} \equiv 0 \nonumber \\
\xi_{A}(T;t) = S \left( m_{RA} \left( T-T_{c,RA} \right)  \right) \nonumber \\
\xi_{R}(T;t) = S \left( -m_{RA} \left( T-T_{c,RA} \right)  \right) \nonumber
\label{eq:LimCase4}
\end{eqnarray}
\end{itemize}
\noindent where $T_{c,ij}$ are the critical temperatures at which the maximum conversion rate with temperature is attained and occurs midpoint between the respective start and finish temperatures of transition, $T_{s,ij}$ and $T_{f,ij}$. The respective $m_{ij}$ are constant parameters determining the steepness of the logistic curve. One can derive an equivalence relationship between the pairs $T_{s,ij}$, $T_{f,ij}$ and $T_{c,ij}$, $m_{ij}$.

It is now straightforward to obtain the functions $F_{i \rightarrow j} \left( T \right)$. For instance,  $F_{M \rightarrow A} \left( T \right)$ is derived by combining \eqref{subeq:RawHeating1} and the first equation in \eqref{eq:LimCase1}, using the identity $1-S(x)=S(-x)$: 

\begin{subequations}
\label{eq:F_functions}
\begin{equation}
 F_{M \rightarrow A} \left( T \right) =m_{MA} a_{MA} S \left( m_{MA} \left( T-T_{c,MA} \right)  \right)
 \label{subeq:F_functionMA}
\end{equation}

Similarly, we obtain the other two functions:
\begin{equation}
 F_{R \rightarrow A} \left( T \right) =m_{RA} a_{RA} S\left( m_{RA} \left( T-T_{c,RA} \right)  \right)
 \label{subeq:F_functionRA}
\end{equation}
\begin{equation}
 F_{A \rightarrow R} \left( T \right) =m_{AR} a_{AR} S \left( -m_{AR} \left( T-T_{c,AR} \right)  \right)
 \label{subeq:F_functionAR}
\end{equation}
\begin{equation}
 F_{R \rightarrow M} \left( T \right) =m_{RM} a_{RM} S \left( -m_{RM} \left( T-T_{c,RM} \right)  \right) 
 \label{subeq:F_functionRM}
\end{equation}
\end{subequations}

\begin{figure}[b]
    \centering
    \begin{tabular}{c}
        \includegraphics[scale=0.5]{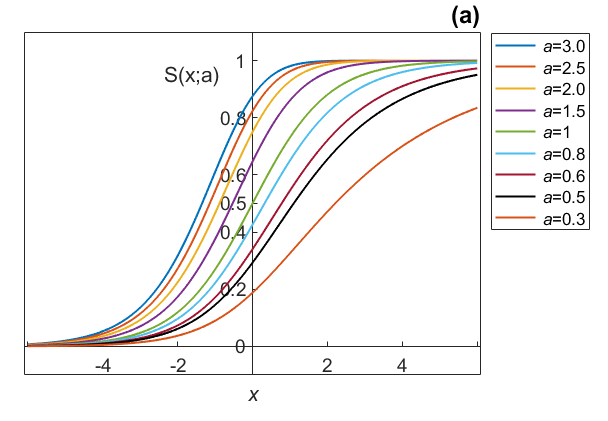} \\
        \includegraphics[scale=0.5]{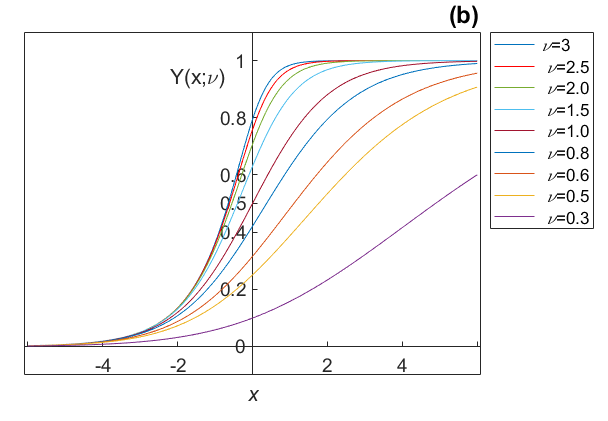}
    \end{tabular}
    \caption{(a) Generalized sigmoidal function $S(x;a)$ for various values of skewness parameter $a$ (b) Richard's function $Y(x;\nu)$ for various values of parameter $\nu$.}
    \label{fig:Sigmoidals} 
\end{figure}

In the equations above, we arbitrarily introduced some extra constant multiplicative factors, $a_{MA}$, $a_{RA}$, $a_{AR}$, $a_{RM}$ in general different for each transition. When equations \eqref{eq:F_functions} are inserted back into the respective differential equations in \eqref{eq:RawHeating} and \eqref{eq:RawCooling} and after integrating these again (considering \eqref{eq:LimCase1} -\eqref{eq:LimCase4}), we derive a new, more general sigmoid function given by:

\begin{eqnarray}
\xi _{i} \left(T \right)= 1 - \left( 1 + e^{m_{ij} \left( T-T_{c,ij} \right)} \right)^{-a_{ij}}, &  \dot{T}>0 \nonumber \\
\xi _{i} \left(T \right)= \left( 1 + e^{m_{ij} \left( T-T_{c,ij} \right)} \right)^{-a_{ij}}, &  \dot{T}<0
\label{eq:GeneralSigmoid}
\end{eqnarray}
\noindent where $\dot{T}$ denotes time derivative of temperature. Therefore, the $a_{ij}$ introduced in the model, have the effect of shifting the maximum of the temperature rate of change of the sigmoid (still occuring at $T=T_{c,ij}$) from  $\xi_i=1/2$ to $\xi_i=1-2^{-a_{ij}}$. If $a_{ij}>0$, the maximum rate occurs at a fraction value of the target phase less than $1/2$ (more than $1/2$ for cooling) and more than $1/2$, if $a_{ij}<0$ (less than $1/2$ for cooling). The $a'$s also affect the steepness of the sigmoidal on one end of the curve (depending on whether they are greater or less than 1) thus breaking the symmetry of the standard logistic curve. We will thus refer to them as "skewness" parameters. For $a_{ij}=1$, \eqref{eq:GeneralSigmoid} reverts to the standard logistic function. The above modified sigmoid resembles somewhat to the so called \textit{generalized} logistic curve, or else, Richard's curve \cite{Richards1959}, given by $Y(x;\nu)=(1+e^{\nu x})^{-\frac{1}{\nu}}$. $Y(x;\nu)$, which also depends on an extra steepness parameter $\nu$, and has similar characteristics with the sigmoid used here. Richard's function has been used before in the description of the martensitic/austenitic transitions in SMAs (for instance see \cite{Zotov2014,kyriakides2012new}). It has also received physical justification by a semi-empirical approach in \cite{Zotov2014}. Nevertheless, we deem that, for the purposes of a dynamic simulation model, the generalized sigmoidal proposed here is more convenient because (i) it emerges by simply inserting a multiplicative factor into equations \eqref{eq:F_functions} and (ii) the final form of the equations \eqref{eq:RawHeating} and \eqref{eq:RawCooling} comes out significantly simpler than if one used Richard's function in equations \eqref{eq:LimCase1} through \eqref{eq:LimCase3} instead. In fig.~\ref{fig:Sigmoidals} we compare the forms of both sigmoids for various values of the "skewness" parameters.

We now state the relationship between the pairs $T_{c,ij}$, $m_{ij}$ and $T_{s,ij}$, $T_{f,ij}$ in accordance to the sigmoid in \eqref{eq:GeneralSigmoid}, assuming that the start temperature occurs when the target phase fraction is $c$ and the finish temperature when the target phase fraction is $1-c$ (c must be small, for instance less than 0.05):

\begin{subequations}
\label{eq:TcMvsTsTf}
\begin{equation}
 m_{ij}=\frac{C_f - C_s}{T_{f,ij} - T_{s,ij}} 
 \label{subeq:m}
\end{equation}
\begin{equation}
 T_{c,ij}=T_{s,ij}-\frac{C_s}{m_{ij}}
 \label{subeq:Tc}
\end{equation}
\end{subequations}
\noindent where $C_s=\ln{\left(1-c\right)^{-1/a}-1}$ and $C_f=\ln{c^{-1/a}-1}$.

Finally, inserting equations \eqref{eq:F_functions} into equations \eqref{eq:RawHeating} and \eqref{eq:RawCooling}, one gets, together with restriction \eqref{eq:KsiRestrictions}, a system of differential equations whose solution gives the general dependence of each lattice fraction on temperature for any values of the four state variables $\xi_{M}, \xi_{A}, \xi_{R}, T$.

In a dynamical situation, the sample temperature varies with time, for example, by self-heating caused by a time-varying current pulse and subsequent cooling by heat advection to the environment and/or radiative heat loss. The temperature variation is obtained by using the heat balance equation. Assuming a time-varying current input signal $I \left( t \right)$ that causes Joule heating, heat advection, and possibly radiative heat exchange with the environment, heat-balance yields:

\begin{equation}
\label{eq:HeatBalance}
 c(T) \cdot d \cdot V \frac{dT}{dt}=I^{2}(t)R(T)-h(T) A \left( T-T_{a} \right) - \sigma  \varepsilon_{T} A \left( T^{4}-T_{a}^{4} \right) - H_{l,MA} \cdot d \cdot V\frac{d\xi_M}{dt}
\end{equation}

where $c_T$ is the specific heat capacity of the alloy that may also depend on temperature (weakly in the temperature ranges at interest), \textit{d} is the density, $V$ its volume, $A$ the total surface area, $h(T)$ the convective heat loss rate to the environment per unit area , $\varepsilon_{T}$ the thermal emmissivity of NiTi (both also temperature dependent), $\sigma$ the Stefan-Boltzmann constant,  $T_{a}$  is the temperature of the environment and  $R \left( T \right) = \rho  \left( T \right) \frac{l}{S}$ is the resistance of the NiTi filament, $l$ being its length and $S$ its cross-sectional area. The last term in \eqref{eq:HeatBalance} represents the latent heat release or absorbed during a transition. However, in the present application of the model we will not include the contribution of latent heat to heat balance, since for the purposes of fitting the available experimental data, latent heat is suppressed due to the small external stress applied to the specimens, as described later on. We also ignore stress/strain related terms in the heat equation, as we did for the equations describing the phase transitions too. We mention here that the applied stress on the filament in the experimental setup to be modeled was rather small. Thus, it is not expected to cause significant change in the sample strain, but only to affect the start and finish temperatures of the transitions, which are considered as free fit parameters in the present model. We also ignored changes in volume or sample length due to thermal effects, as these are less than 1\% in the temperature ranges considered. For a sufficiently thin filament, a single uniform temperature value for the entire sample is a good approximation. Regarding the temperature dependence of $c(T)$ and $h(T)$, we devote a subsection later on in the paper. $\varepsilon(T)$, the NiTi emmissivity, has also been shown to depend very weakly on temperature (e.g in \cite{daSilva2016}), but also to obtain rather different values between cooling and heating. For temperatures below $\approx 200 ^\circ$C, the radiative term is very small compared to the conduction term (assuming experiments are not conducted in vacuum), thus we ignore the temperature dependence of emissivity, but, for consistency with available experimental data, we will use a single constant value, one for heating, $\varepsilon_{heat}(T)$, and one for cooling, $\varepsilon_{cool}(T)$. The values used were taken from \cite{daSilva2016} by using the reported values of the "direct" measurements at a temperature around 60~$^\circ$C, which roughly correspond to the average values for emmissivity of the heating and cooling cycles over the entire temperature range considered therein.

It is convenient to rewrite equation \eqref{eq:HeatBalance} in dimensionless form by appropriately rescaling time $t$:

\begin{equation}
 c'(T') \frac{dT'}{dt^{'}} = P_{o}I^{'2}(t') \rho^{'} (T') - h'(T') \left( T'-1 \right) - \mu^X_{\varepsilon} \varepsilon'(T') \left( T^{'4}-1 \right),
 \label{eq:HeatBalanceDimless}
\end{equation}

\noindent where primed quantities are the dimensionless counterparts of time $t$, current $I$, resistivity $\rho$ and temperature $T$ respectfully, defined by the following change of variables:

\begin{subequations}
\begin{equation}
t=t_{o}t^{'}, ~ t_{o}=\frac{c_a \cdot d \cdot S}{\Pi h_a}
\label{subeq:Dimensionless_t}
\end{equation}
\begin{equation}
 I=I_{o}I^{'}%, \nonumber 
\label{subeq:Dimensionless_I}
\end{equation}
\begin{equation}
 \rho = \rho_{aM} \rho'%, \nonumber 
\label{subeq:Dimensionless_rho}
\end{equation}
\begin{equation}
T=T^{'}T_{a}
\label{subeq:Dimensionless_T}
\end{equation}
\label{eq:HeatBalanceParams}
\end{subequations}

\noindent In the above equations $I_{o}$ is the maximum current value in the input current signal, $\rho_{aM}$ is the resistivity of M-phase at ambient temperature, $\Pi$ is the \textit{perimeter} of the filament, assumed constant across its entire length, and 

\begin{eqnarray}
\label{eq:DimensionlessQuantities}
c'(T') = c(T)/C_0, \nonumber \\
P_{o}=\frac{I_{o}^{2} \rho_{aM}}{h_a S \Pi T_{a}} \nonumber\\
h'(T')=\frac{h(T')}{h_a} \nonumber\\
\mu^X_{\varepsilon}(T) =\frac{ \sigma \varepsilon_X(T') T_{a}^{3}}{h_a} \nonumber\\
T'_{c,ij}=\frac{T_{c,ij}}{T_a}, m'_{ij} = m_{ij} T_a,  i,j = 1 \ldots N, i \neq j
\end{eqnarray}

\noindent In the above, $h_a$ is the heat transfer coefficient at ambient temperature and $C_0$ a reference constant specific heat capacity (see the next subsection). Replacing the derivatives with respect to temperature by derivatives with respect to time in system \eqref{eq:RawHeating}, \eqref{eq:RawCooling}:  $\frac{d \xi _{i}}{dT^{'}}=\frac{d \xi _{i}}{dt^{'}} \cdot \frac{dt^{'}}{dT^{'}}= \dot{\xi_{i}} \cdot  (\dot{T})^{-1}$, we finally obtain the following system of equations for the time-dependent variables $\xi_{i}(t')$ and temperature $T'$:

\begin{subequations}
\label{eq:whole}
\begin{equation}
    \label{subeq:1}
 \dot{\xi}_{M}= \left\{
 \begin{matrix}
- \xi_{M}F_{M \rightarrow A}(T^{'}) \dot{T^{'}},~ \dot{T^{'}}>0\\
\\
\xi_{R} F_{R \rightarrow M}(T^{'}) \dot{T^{'}},~ \dot{T^{'}}<0\\
\end{matrix}
\right.
\end{equation}

\begin{equation}
    \label{subeq:2}
\dot{\xi}_{A} = \left\{ \begin{matrix}
 \left[  \xi_{M} F_{M \rightarrow A}( T^{'}) + \xi_{R} F_{R \rightarrow A}( T^{'})  \right] \dot{T^{'}}, \dot{T^{'}}>0\\
\\
 - \xi _{A} F_{A \rightarrow R}( T^{'}) \dot{T^{'}},~ \dot{T^{'}}<0\\
\end{matrix}
\right.
\end{equation}

\begin{equation}
    \label{subeq:3}
\xi _{R} = 1 - \xi _{M} - \xi _{A}
\end{equation} 

 \begin{equation}
    \label{subeq:4}
 c'(T') \dot{T^{'}}=P_{o}i^{'2}(t^{'}) \rho^{'}(T^{'}) - h'(T') \left( T^{'}-1 \right) - \mu_x \left( T^{'4}-1 \right)
\end{equation}

\end{subequations}
\noindent where the subscript "x" stands for either "heating" or "cooling". In all of equations \eqref{eq:whole}, the dotted quantities are time derivatives. In \eqref{subeq:4}, $\rho^{'} \left( T^{'};t' \right)$ is calculated from \eqref{eq:SampleResistivity} by inserting the re-scaled quantities:

\begin{equation}
\label{eq:ResVsTDimless}
 \rho^{'}(T^{'};t^{'}) = \xi _{M}(t^{'})  \rho^{'}_{M} (T^{'}) + \xi_{A}(t) \rho^{'}_{A}(T^{'}) + \xi _{R}(t^{'})  \rho^{'}_{R}(T^{'})
\end{equation}

\noindent where:

\begin{subequations}
  \begin{equation}
  \rho'_M(T')=1+a'_M(T'-1), \\ 
\label{eq:ResvsTDependence_M} 
\end{equation}
  \begin{equation}
  \rho'_A(T') = \beta_{AM} \left[ 1+a'_A (T'-1)  \right], \\
\label{eq:ResvsTDependence_A} 
\end{equation}
  \begin{equation}
  \rho'_R(T') = \beta_{RM} \left[ 1+a'_R (T'-1)  \right],
\label{eq:ResvsTDependence_R} 
\end{equation}
  \label{eq:ResvsTDependence}  
\end{subequations}

\noindent are the dimensionless counterparts of equations \eqref{eq:ResVsT}. The model parameters $\beta_{AM}$ and $\beta_{RM}$ are the respective ratios of A-phase to M-phase and R-phase to M-phase resistivities at ambient temperature and  $a'_M$, $a'_A$, $a'_R$ are respectively equal to $a_M$, $a_A$, $a_R$ multiplied by $T_a$. In \eqref{eq:whole}, $T$ is replaced by $T'$ and similarly $T_{c,MA}$, $T_{c,AR}$, $T_{c,RM}$ are replaced by $T^{'}_{c,MA}$, $T'_{c,AR}$, $T'_{c,RM}$, each corresponding to the ratio of the respective critical temperature to the ambient temperature. Finally, the slopes $m_{ij}$ are replaced by their dimensionless counterparts: $m'_{ij}=m_{ij} T_a$.
 
The system of equations \eqref{subeq:1}, \eqref{subeq:2} and \eqref{subeq:4} is a non-autonomous system of first-order differential equations, driven by a time-dependent input current signal and accompanied by the algebraic restriction \eqref{subeq:3}. For any set of initial conditions of the four dynamic variables $\xi_{A}$, $\xi_{M}$, $\xi_{R}$ and $T^{'}$ the system can be integrated numerically. Combined with equations \eqref{eq:ResVsTDimless} and \eqref{eq:ResvsTDependence}, it yields the dynamical evolution of the sample resistivity $\rho^{'}$ and voltage $V(t)=I(t) R(t)$.

\subsection{\label{subsec:convectiveheat} Convective heat transfer coefficient and specific heat capacity temperature dependence}

Assuming a long cylindrical NiTi wire placed in air, $h(T)$ can be written as:

\begin{equation}
h\left( T \right) = \frac{k(T)}{D} N_u(T),
\label{eq:hRaw}
\end{equation}

where $k$ is the thermal conductivity of air, $D$ the diameter of the wire and $N_u$ the Nusselt number which can be related to the Rayleigh number $Ra$ with the empirical formula 

\begin{equation}
Nu = A + C Ra^m
\label{eq:Nusselt}
\end{equation}

\noindent where $A, C, m$ are constants that depend also on the orientation of the wire relative to the horizontal. Rayleigh number can be written as the product of the Prandtle number $Pr$ and the Grashof number $Gr$ which are given by:

\begin{eqnarray}
Ra  = Pr \cdot Gr, \nonumber \\
Pr=\frac{\mu C_p}{k} \nonumber \\
Gr=\frac{2 \left( T - T_0\right) g D^3}{\nu^2 \left( T + T_0\right)}
\label{eq:Rayleigh}
\end{eqnarray}

\noindent where $g$ is the acceleration of gravity, $C_p, \mu, \nu$ are respectfully the specific heat capacity, kinematic viscosity and viscosity \textit{of surrounding air}, all evaluated at some average temperature of the range considered for $h(T)$. $T_0$ is the room temperature for which the correlation parameters $A$, $C$, $m$ have been experimentally determined. Combining \eqref{eq:hRaw}, \eqref{eq:Nusselt} and \eqref{eq:Rayleigh}, we finally obtain:

\begin{equation}
h'\left( T' \right) = \left[ \frac{A k}{D} + B D^{3 m - 1} \left( \frac{T' - T_0/T_a}{T' + T_0/T_a} \right)^m \right]/{h(T_a)},
\label{eq:hSpecific}
\end{equation}

\noindent We wrote \eqref{eq:hSpecific} directly in dimensionless form and as a function of dimensionless temperature $T'$. Also, $B = C k^{1-m} \left( \frac{2 \mu g C_p T_a}{\nu^2} \right)^m$.
For the values of the involved physical parameters of air we used the values presented by Talebi \textit{et al.} in \cite{TaleHosse14} (see Table~2 therein). For the correlation parameters $A$,~$C$ and $m$ we used the parameters reported in \cite{Eisa2011} for small external stress, an horizontal wire, $T_0=293~^\circ$K and a NiTi sample by Dynalloy Inc. (CA, USA) of roughly the same diameter as the sample used for the resistivity measurements used in the present study.

\begin{figure}[b]
    \centering
        \includegraphics[scale=0.70]{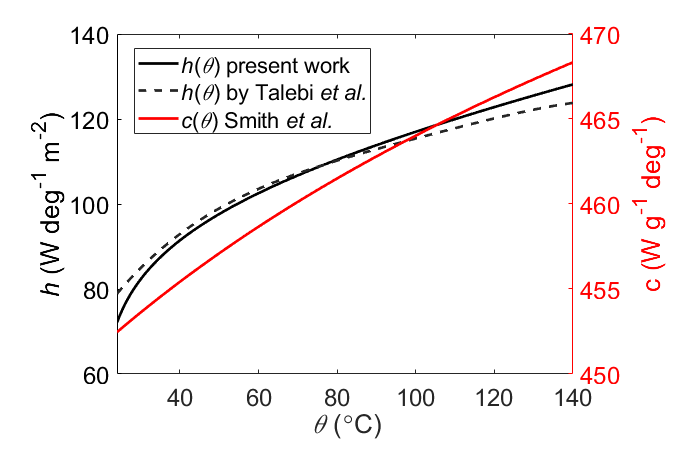}
    \caption{Temperature dependence of the heat transfer coefficient (left vertical axis) and the specific heat capacity (right vertical axis) used in the present study. For comparison, we show the temperature variation of heat transfer coefficient used by Talebi \textit{et al.} in \cite{TaleHosse14}.}
    \label{fig:handCvsTemperature} 
\end{figure}

Regarding the temperature variation of specific heat, we used an empirical formula, a fit to experimental data for NiTi, as reported by Smith \textit{et. al.} \cite{Smith1993}:

\begin{equation}
c(T) = C_0 + \gamma_1 T - \frac{\gamma_2}{T^2},
\label{eq:CSpecific}
\end{equation}
\noindent with $C_0=0.46395$~J$\cdot$deg$^{-1}$/kg, $\gamma_1=3.982\cdot10^{-5}$~J/kg and  $\gamma_2=2.06\cdot10^3$~$\cdot$deg/kg. Equation \eqref{eq:CSpecific} is valid for the high-temperature region of the alloy. 

In fig.~\ref{fig:handCvsTemperature} we plot the temperature variation of the heat transfer coefficient and the specific heat capacity as calculated from the equations above. For comparison we show the temperature variation of heat transfer coefficient used by Talebi \textit{et al.} \cite{TaleHosse14} for similar experimental conditions to ones used in the present work.

\section{\label{sec:SimRes}Experiments and Model Validation}

In this section we perform numerical simulations and compare them to available measurements published elsewhere \cite{stavrinides2019niti}. We use two kinds of experimental resistivity data: (i) \textit{Passive-heating data}: This corresponds to electrical measurements on a NiTi thin filament passively heated/cooled vy varying the temperature of the environment at small temperature increments allowing the system to reach thermal equilibrium after each change. (ii) \textit{Self-heating data}: This comes from a set of measurements performed by applying periodic time-varying current pulses in the frequency range of $f=0.01-10$~Hz in order to capture the dynamics of the resistivity variation. For the numerical calculations, we integrated the dimensionless system (\ref{eq:whole}) using MATLAB ODE15s solver by using the same periodic current signal and the frequency values of the experimental studies. Finally, the results from the simulations and the measurements are compared.

\subsection{Experimental samples and measurement data}

As already mentioned, experimental data was taken from \cite{stavrinides2019niti}, an experimental study that used off-the-shelf NiTi wires available from Dynalloy, Inc company under the trademark Flexinol\textsuperscript{\texttrademark}. In the said work, NiTi samples had the form of bare, crimped wires, almost equiatomic (Ni/Ti ratio was 51:49), measuring 200 $\mu$m in diameter, and lengths of approximately 15~cm. Martensite to Austenite transition was reported to be almost completed at $\approx 90^\circ$C. Two kinds of experiments were performed there: (i) \textit{passive heating experiment}: the samples were placed in an oil-bath (on a Peltier element) and external temperatures were slowly varied in small steps from $0^\circ$C to $91^\circ$C (heating round) and $91^\circ$C to $0^\circ$C (cooling round), allowing the sample to remain at the same temperature for several minutes so that thermal equilibrium is established. After each temperature change, four-probe resistance measurements were performed using an Agilent 34401A DMM. In order to let the sample lattice structure relax, a few heating-cooling rounds were performed using a freshly cut specimen, and then resistance measurements were recorded for the last heating-cooling round. (ii) \textit{self-heating experiments}: freshly cut specimens were self-heated in air at room temperature ($\approx 24^\circ$C) by issuing a ramp (triangular) periodic current signal of various frequencies produced by a voltage-controlled current source. Resistance at each time instance was determined by a measurement of the voltage across the sample and the value of the current at that instance. The above study reported resistance \textit{vs.} current curves as well as the I-U characteristics for each driving signal frequency.

\subsection{\label{subsec:SimulationMethodology}Methodology of numerical simulations}

\subsubsection{\label{subsubsec:FixedParams} Assigning values to fixed parameters}
For determining model parameters that best fit the available experimental data, we first fixed the values of the thermal properties of the NiTi filament entering the heat-balance equation, \textit{i.e.} the constant coefficients of the specific heat in \eqref{eq:CSpecific}, the heat transfer coefficient function $h(T)$ in \eqref{eq:hSpecific} and $\varepsilon_{heat}$, $\varepsilon_{cool}$. We also fixed the constant geometrical characteristics of the filament entering the model ($A$, $S$), the maximum current value $I_0$ and frequency $f$ of the linear-ramp current pulses, as are all reported in \cite{stavrinides2019niti}. The ambient temperature $T_a$ was set equal to the known ambient temperature of the measurements ($T_a\approx24$~$^\circ$C). In order to better use the noisy experimental resistivity \textit{vs.} current data, we applied a simple running average smoothing of $\rho$ \textit{vs.} $I$ for a smoothing window of 200 data steps for the $f=0.01$~Hz measurements. The resulting smoothed curve is depicted in \ref{fig:rhovsIsimulation}a. The noise in experimental data of all the pulse measurements comes mainly from the large propagated errors of the calculation of $R(t)$ from $V(t)/I(t)$, especially at values of $I(t)$ close to zero. The value of resistivity of A-phase at 90~$^\circ$C is thus set equal to the resistance value of the sample occurring at a current value $I(t)\approx0.45$~A (heating round), where the austenitic transition seems to be completed and the linear resistance region begins. For the temperature dependence of A-phase resistivity at the linear region (\textit{i.e.} for temperatures higher than $T_{f,MA}$ and $T_{f,RA}$) corresponding to $I>0.45$~A, we used the slope of a linear fit to $\rho$ \textit{vs.} $I$ data in the current range $0.45-0.5$~A, as shown in the figure inlet, and thus fixed $a'_A$. We then linearly extrapolated the values of the single A-phase resistivity $\rho_A(T)$ (assuming linear dependence on temperature) for all other temperatures. For the values and temperature dependence of $\rho_M$ and $\rho_R$ we used data found in literature for Flexinol NiTi (49/51) \cite{NOVAK2008127}. Because they do not exactly match the values of the resistivity for the A-phase obtained from the experimental data of \cite{stavrinides2019niti} used here, we adjusted $\rho_M$ and $\rho_R$ so that their respective ratios to $\rho_A$ are the same as the ratios of the reported data in \cite{NOVAK2008127}. We should note that, although one it was possible to determine $\rho_A(T)$ at the linear region from the high-current available data, it is was \textit{not} possible to do the same for $\rho_M(T)$ and $\rho_R(T)$ from the low-current data, because data is too noisy there. Neither was it possible to do so from the low-temperature $\rho$ \textit{vs.} $T$ experimental data obtained from the passive-heating experimental data (shown in figure \ref{fig:res_vs_temp}), because at low temperatures the lattice is expected to be at a mixed state of M- and R-phase, thus resistivity of each individual phase cannot be directly inferred. An additional complication comes from the fact that the passive-heating measurements in \cite{stavrinides2019niti} were recorded after several heating-cooling rounds, in order to allow the lattice structure to relax, whereas the self-heating measurements at the low frequency $f=0.01$~Hz were performed with fresh samples and recording started from the first round. Thus a different M-, R-phase mixture is expected between the two, leading to different resistivity values and different temperature dependence. In summary, the values of all the fixed parameters used are shown in table~\ref{tab:FixedParamsTable}.

\subsubsection{\label{subsubsec:FreeParams} Model free parameters \& simulation procedure}

The free model parameters are: (i) the transition temperatures $T_{c,MA}$, $T_{c,RA}$, $T_{c,AR}$, $T_{c,RM}$, (ii) the sigmoid slopes $m_{MA}$, $m_{RA}$, $m_{AR}$ and $m_{RM}$, (iii) the sigmoid "skewness" parameters $a_{MA}$, $a_{RA}$, $a_{AR}$ and $a_{RM}$ and (iv) the \textit{initial} fractions $\xi_M(0)$, $\xi_A(0)$ and $\xi_R(0)$ of each phase ($t=0$).

We assigned non-zero initial values to fractions $\xi_M(0)$ and $\xi_R(0)$ assuming that a fresh sample contains zero A-phase proportion at low temperatures and is just a mixture of M-, R-phase. Having fixed the resistivities and their temperature dependence of the two phases as discussed in the previous section, one can roughly infer the initial ratio of M-phase relative to R-phase from the measured resistivity values at the low current region using the smoothed data of the $f=0.01$~Hz experiments. In this experiment three heating-cooling cycles are recorded as shown in \ref{fig:rhovsIsimulation}a (raw data) and \ref{fig:Res_vs_I_f=0.01_smoothed} (smoothed data) \footnote{Only the last cycle of the $f=0.01$~Hz measurements was originally presented in \cite{stavrinides2019niti}. Available unpublished data for $f=0.01$~Hz was provided by the authors of that work for the purposes of this paper}. Unfortunately, resistance data from the 1st heating cycle is only available in part, the first half of the 1st heating cycle not having been recorded, the second half can be seen in the figures. By adjusting the rest of the free parameters, we performed numerical simulations until the system reaches a steady state cycle for each one of four different frequency values, $f=0.01, 0.1, 0.5, 10.0$~Hz. The fine-tuning of the free parameters involved concurrently considering the lowest ($f=0.01$~Hz), medium ($f=0.1$,$0.5$~Hz) and high ($f=10$~Hz) frequencies as each one data set helps raise much of the degeneracy in the values of the start and finish temperatures of each pair of transitions: more than one sets of transition temperatures for all the allowed transitions are able to fit perfectly the lowest frequency data, but most of them will perform very poorly on the medium and highest frequencies. There seems to be an optimum set of free parameter values, the right one in physical terms, that matches sufficiently well the resistance measurements \textit{vs.} current for all frequencies. It should be noted that, for the self-heating experiments, where temperature is dynamically varying, the transition temperatures are very sensitive to external conditions, i.e. ambient temperature and especially the heat transfer coefficient. Even small changes in these can cause significant changes in the temporal evolution of resistivity data. The resistivity variation among individual lattice phases is small anyway, as the phenomenon studied involves small-scale changes. Therefore, one should be very careful at determining the precise heat exchange conditions when comparing experiments and simulation under rapidly changing temperatures.   

After determining the optimal parameters for fitting the self-heating experiments, we performed numerical simulations for passively heating experiments. Passive heating experiments were conducted at different external conditions from self-heating experiments in \cite{stavrinides2019niti}. First of all, there was no external stress applied, therefore the transition temperatures are expected at lower values than in the self-heating experiments. Secondly, under slow passive heating heat conduction to environment is not important for the outcomes. Also, there is no Joule heating and the temperature rise is caused only by changing the temperature of the environment until the sample temperature equilibrates. For the simulations, we started with the exact same fixed and free parameter values as we the self-heating runs, but now we zeroed the Joule heating term and varied the ambient temperature $T_a$ from 0 to 92~$^\circ$C and back to 0, in increments of 1~deg, allowing enough time to pass for the sample to reach equilibrium temperature after each change. We performed simulations assuming a fresh NiTi sample with the exact same initial proportions $\xi_M(0)$, $\xi_A(0)$ and $\xi_R(0)$, as in the self-heating simulations. We performed several heating-cooling cycles so that we capture the dynamical relaxation of the filament's lattice structure, thus reproducing the experimental conditions of the particular experiment in \cite{stavrinides2019niti}. 

\begin{table}[th]
    \centering
    \begin{tabular}{c|c||c|c||c|c}
    Parameter & Value & Parameter & Value & Parameter & Value\\ \hline
$I_0$ &	0.5 A &	$T_{a}$ &	24 $^\circ$C & $D$ & $6.450$ g/cm$^3$ \\
$\rho_M(10~^\circ$C) &	$7.273\times10^{-7} \Omega m$ &	$\rho_A(10~^\circ$C) &	$8.000\times10^{-7} \Omega m$ &	$\rho_R(10~^\circ$C) &	$9.939\times10^{-7} \Omega m$\\
$a_{M}$ &	$8.250\times10^{-4}$ deg$^{-1}$& $a_{A}$ &	$2.562\times10^{-4}$ deg$^{-1}$&
$a_{R}$ &	$2.515\times10^{-4}$ deg$^{-1}$ \\
$\varepsilon_{heat}$ & 0.170 & $\varepsilon_{cool}$ & 0.202 & - & - 
    \end{tabular}
    \caption{Values of model fixed parameters.}
    \label{tab:FixedParamsTable}
\end{table}

\subsection{\label{subsec:Comparison} Comparison between numerical results and measurements}

The optimal set of values of the free parameters for the self-heating data for all frequencies are presented in table \ref{tab:FreeParameterTable}, where we report the start and finish temperatures $T_{s,ij}$, $T_{f,ij}$ and the corresponding critical temperature $T_{c,ij}$ of the model. Regarding the skewness parameters $a_{ij}$, we found that the values of $a=0.5$ are best for describing the shapes of all transitions. 

\begin{table}[th]
    \centering
    \begin{tabular}{c|c||c|c||c|c}
    Parameter & Value & Parameter & Value & Parameter & Value \\ \hline
$\xi_M(t=0)$ &	0.358 &	$\xi_R(t=0)$ &	0.642 &	$\xi_A(t=0)$ &	0.000\\
$T_{s,MA}$ &	82.45 $^\circ$C &	$T_{f,MA}$ &	127.0 $^\circ$C	& $T_{c,MA}$ &	94.54 $^\circ$C \\
$T_{s,RA}$ &    44.69 $^\circ$C &   $T_{f,RA}$ &	131.30 $^\circ$C &   $T_{c,RA}$ &	68.15 $^\circ$C	\\
$T_{s,AR}$ &	152.55 $^\circ$C &	$T_{f,AR}$ &	36.98 $^\circ$C &	$T_{c,AR}$ &	68.29 $^\circ$C	\\
$T_{s,RM}$ &   123.57 $^\circ$C &   $T_{f,RM}$ &	-17.64 $^\circ$C &   $T_{c,RM}$ &	20.615 $^\circ$C 	\\
$a_{MA}$ &	0.5  &	$a_{RA}$ &	0.5 &  $a_{AR}$ &	0.5  \\
$a_{RM}$ &	0.5 & - & - & - & - \\
    \end{tabular}
    \caption{Fitted values of the model free parameters for the self-heating simulations. Start and finish temperatures are calculated from \eqref{eq:TcMvsTsTf} for $c=0.05$.}
    \label{tab:FreeParameterTable}
\end{table}

\begin{figure}
    \centering
    \begin{tabular}{cc}
        \includegraphics[scale=0.33]{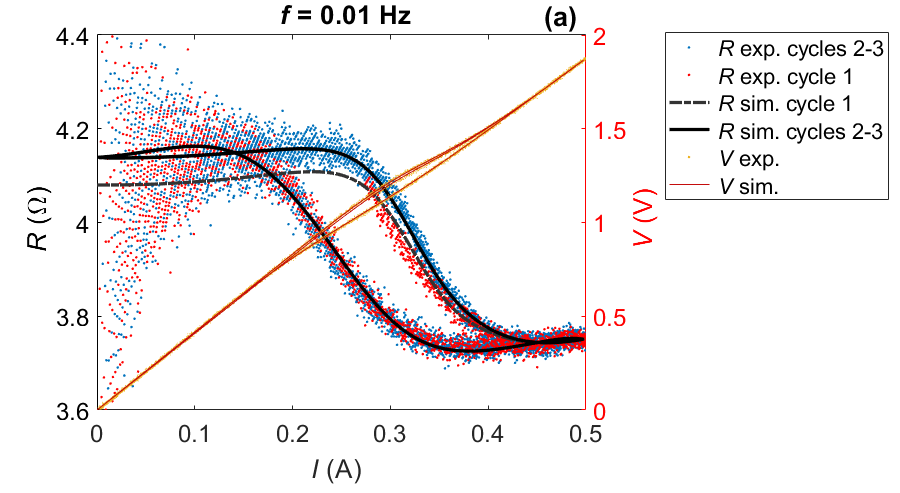} &
         \includegraphics[scale=0.35]{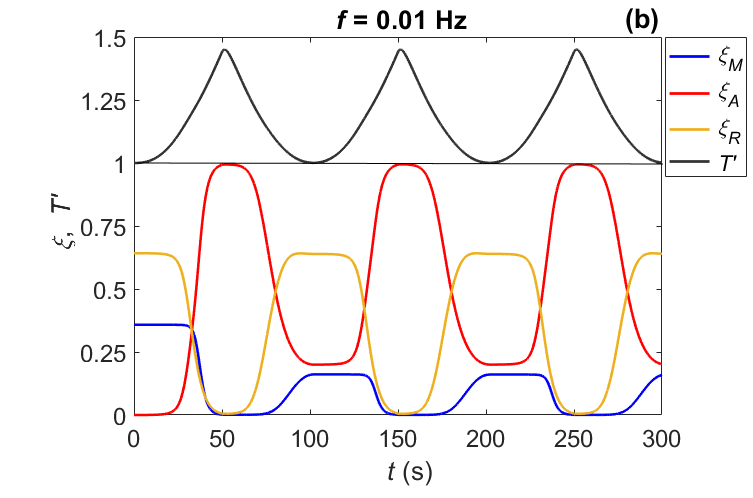} \\

        \includegraphics[scale=0.33]{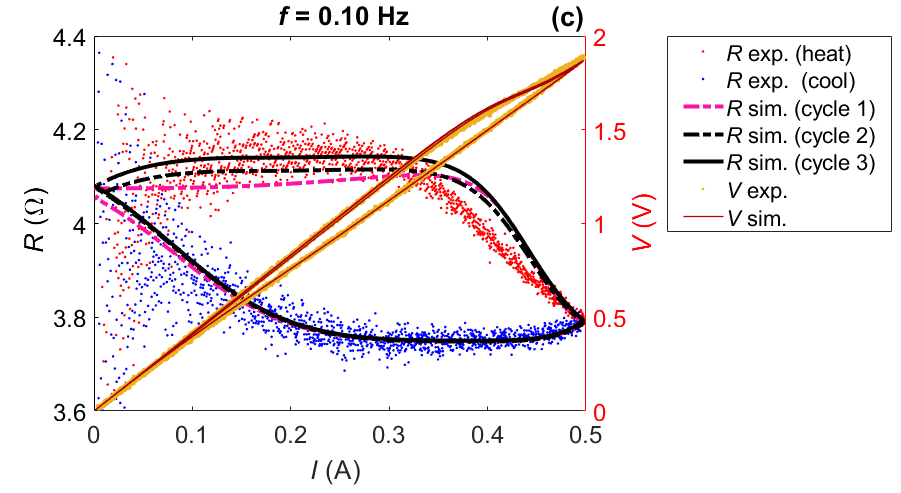} &
        \includegraphics[scale=0.33]{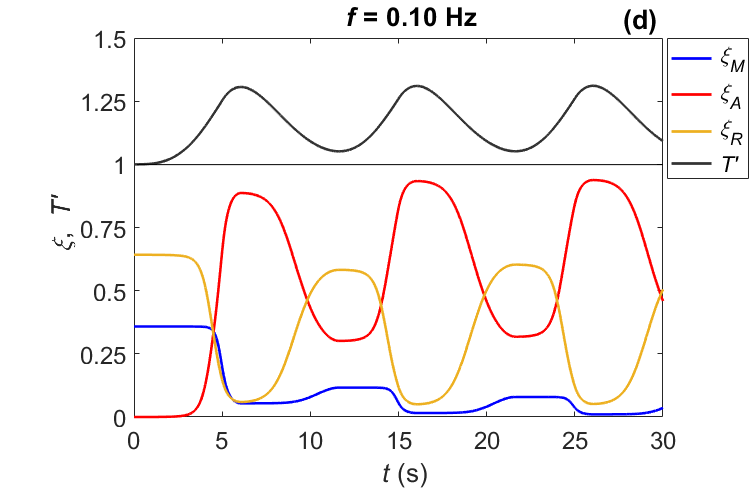} \\

        \includegraphics[scale=0.33]{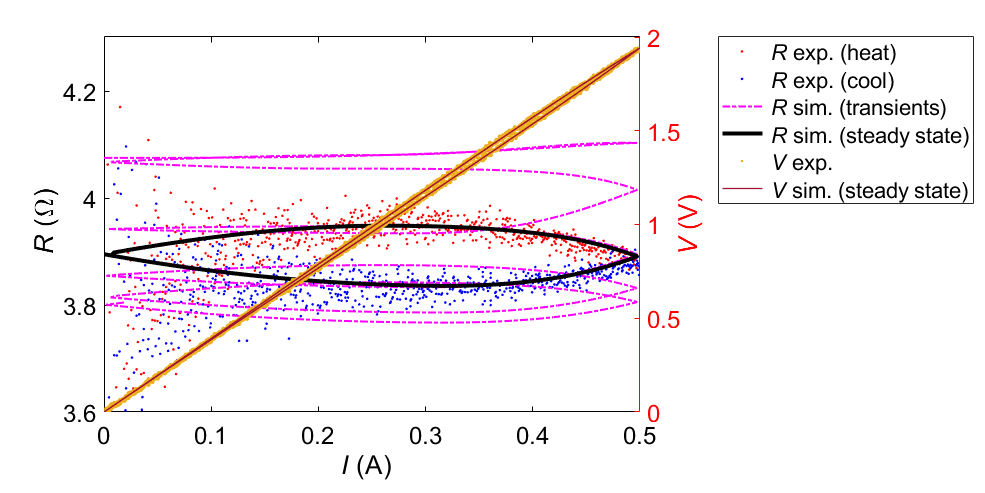} &
        \includegraphics[scale=0.33]{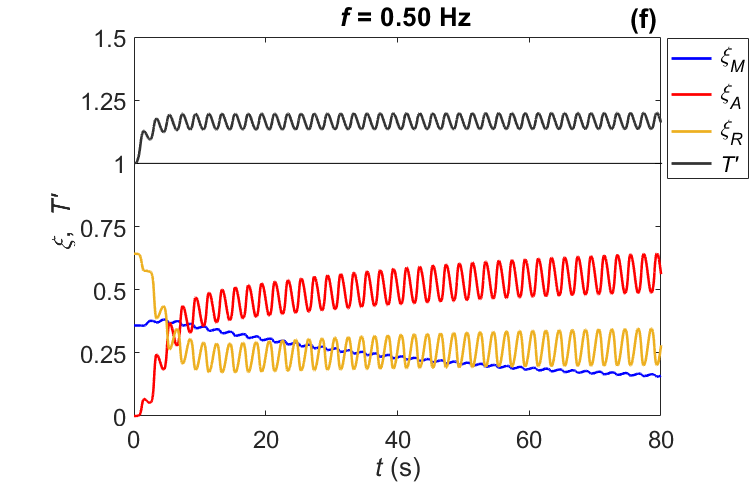} \\
        
        \includegraphics[scale=0.33]{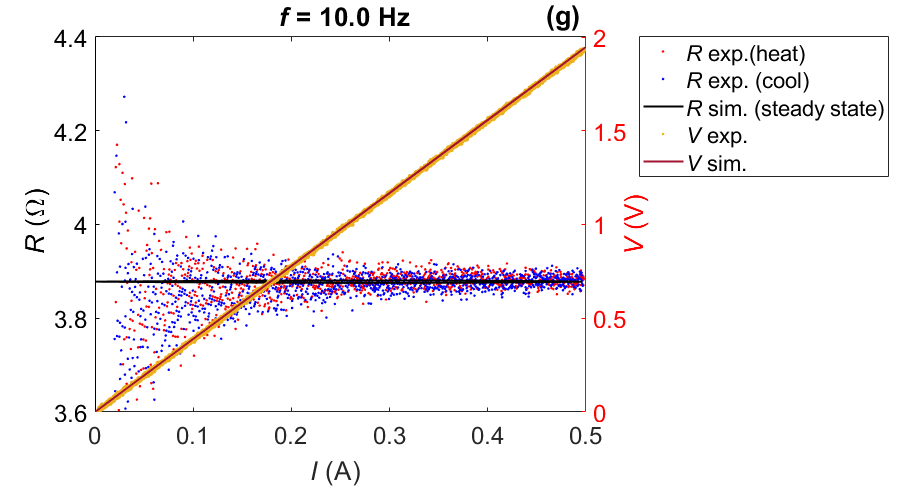} &
        \includegraphics[scale=0.33]{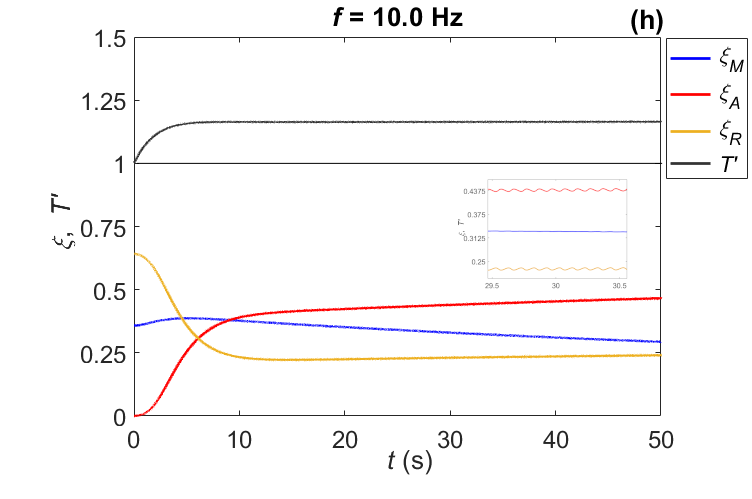}
    \end{tabular}
    \caption{Comparison between numerical results and experimental data for the self-heating experiments for a periodic triangular input current pulses of various frequencies $f$. Plots in the left column show resistance $R vs.$ current intensity $I$ (left vertical axis) and voltage \textit{V vs. I} (right vertical axis). Plots in the right column show the temporal variation of NiTi structural phase fractions $\xi_M, \xi_A, \xi_R$ and the dimensionless temperature $T'$ produced by the model. Simulation results for several heat-cooling cycles are shown for each frequency.
    }
    \label{fig:rhovsIsimulation}
\end{figure}

The respective plots comparing simulation and experimental data for the self-heating experiments are shown in fig.~\ref{fig:rhovsIsimulation}a,c,e,g. In the figures on the left column of fig.~\ref{fig:rhovsIsimulation}, we plot sample resistance \textit{vs.} the driving current ($R$ vs. $I$) on the left vertical axis and the voltage across the sample vs. the current ($V$ vs $I$) on the right vertical axis for the four frequencies considered. Experimental values are denoted by dots (red for heating round, blue for cooling round), while numerical results are depicted by solid curves. In the figures on the right (Fig.~\ref{fig:rhovsIsimulation}b,d,f,h) we show the respective temporal evolution of the fraction of each of the three lattice phases $\xi_M,\xi_A,\xi_R$, and the (dimensionless) temperature $T'$ as predicted by the model. For two of the driving frequencies, 0.01~Hz and 0.1~Hz, three and five heating-cooling cycles are shown respectfully. For $f=0.01$~Hz, the steady-state cycle is almost perfectly reached after only one cycle, while for $f=0.1$~Hz the steady state is reached after 2-3 cycles. For $0.5$~Hz and $10$~Hz, 50 and 500 cycles are shown respectively, for which the steady-state cycle is reached very slowly. In fact, for $f=10$~Hz the resistance equilibrates after more than 5000 cycles (the steady state depicted in \ref{fig:rhovsIsimulation}g is obtained after 8,000 cycles), as the fractions of all phases slowly change despite the fact that the temperature appears constant at long time scale. (Transient curves for the $f=10$~Hz numerical results are not shown in \ref{fig:rhovsIsimulation}g in order for the steady state cycle to appear clearly). The slowly evolving lattice phase fractions at the highest frequency can be attributed to the rapid temperature changes taking place at very short time scales (see blown-up region shown at the figure \ref{fig:rhovsIsimulation}h inlet) leading to a continuous and gradual lattice relaxation occurring over a long period of time. In the $R$ \textit{vs.} $I$ graph of the measurements for $f=0.01$, the apparent difference between cycle~1 and cycles~2,~3 proves that the transition from M-phase to A-phase during heating has different start and finish temperatures from the R- to A-phase transition. In fact, one can determine rather accurately the start and finish temperatures of M- to A- transition by trying to match the \textit{average} level of electrical resistance of the medium and high frequency runs ($f=0.5$ and $f=10$~Hz). The precise values of these critical temperatures affect the relative proportion of M-phase at the steady state, which changes the average resistance level. Equally important are the start and finish temperatures of the R- to M- transition occurring during cooling, at the same time that A-phase is transforming to R-phase. These temperatures determine the level of resistivity rise due to the increase of the proportion of the R-phase, since M-phase has a lower resistivity value than R-phase.

As can be seen in Fig.~\ref{fig:rhovsIsimulation}b, after the 1st heating-cooling cycle, there is non-zero remaining R-phase and even a non-zero A-phase fraction in the sample . This means that during cooling not all the A-phase and R-phase have been converted back to the M-phase, as the temperature does not drop sufficiently in order for the A-to-R and R-to-M phase transitions to be completed. On the other hand, during each heating-cooling cycle, the average proportion of M-phase is gradually reduced while the average proportion of R-phase increases. The existence of R-phase at low values of $I$ is the reason why $R$ is slightly higher there after the 1st cycle. This is in accordance with the experimental $R$ \textit{vs.} $\theta$ data from \cite{stavrinides2019niti} and elsewhere (\cite{bhargaw2013thermo}).

\begin{figure}[b]
    \centering
        \includegraphics[scale=0.70     ]{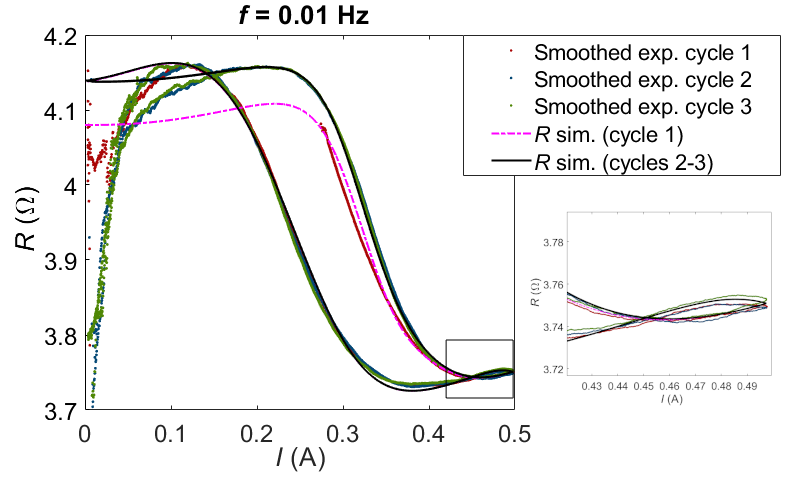}
    \caption{Resistance $vs$.current: comparison between simulation model and experimental results for the \textit{smoothed} experimental data of fig. \ref{fig:rhovsIsimulation}a. The experimental data of each one of the three heating-cooling cycles is depicted with different color.}
    \label{fig:Res_vs_I_f=0.01_smoothed} 
\end{figure}

As already mentioned, in figure \ref{fig:Res_vs_I_f=0.01_smoothed} we present the smoothed raw data of \ref{fig:rhovsIsimulation}a, where we denote the experimental data of each heating-cooling cycle by dots of a different color. Numerical results are overlaid (bold black solid curve) and fit almost perfectly the smoothed data. The inlet figure contains a blow-up of the region prescribed by the rectangular box depicting a 'micro-scale' pinched hysteresis loop evident both in the smoothed experimental data as well as in the simulation results. This loop is caused by the fact that, after the current reaches its peak value and begins to drop, there is a brief period of time where temperature continues to increase causing a linear rise in the resistivity of the pure A-phase filament and then, when the sample begins to cool, resistivity drops again causing the pinch of the curve. The precise location of the pinch as well as the vertical height of the loop depend on the thermal coefficients of A-phase resistivity and the rate of heat loss. On the other hand, the relaxation of the lattice is evident between consecutive cycles, even for the lowest frequency runs. It should be mentioned, that although the agreement between numerical results and measurements is remarkable for both the low frequency and higher frequency data (for the exact same set of fixed and free parameter values), there is some notable difference between simulation and experiment in the $f=0.1$~Hz heating cycle data (see fig.~\ref{fig:rhovsIsimulation}c). It appears as if the start temperature for the R- to A-phase transition predicted by the model occurs at a slightly lower value than the one inferred by the experimental data for that frequency. We tried to amend this discrepancy by several attempts to alter the fit parameters, including changing the skewness parameter $a$ for the involved transitions, but we concluded that there was no single set of model parameters that simultaneously fits perfectly all the other frequencies and the heating cycle of the $f=0.1$~Hz data. We comment a bit further on this issue in the discussion section next.   

\begin{figure}[b]
    \centering
    \begin{tabular}{cc}
        \includegraphics[scale=0.35]{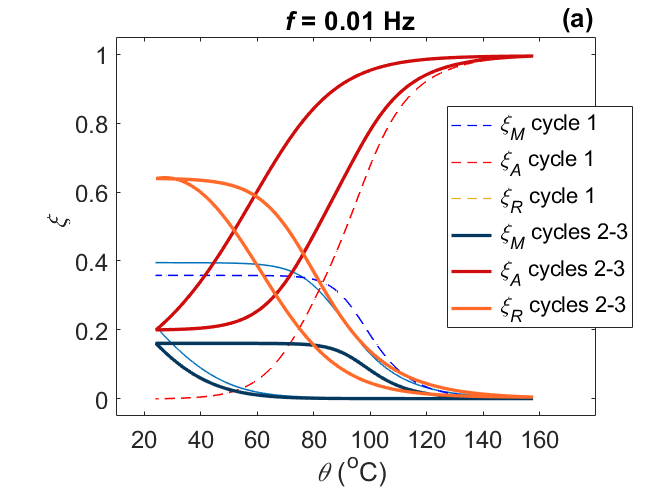} &
         \includegraphics[scale=0.35]{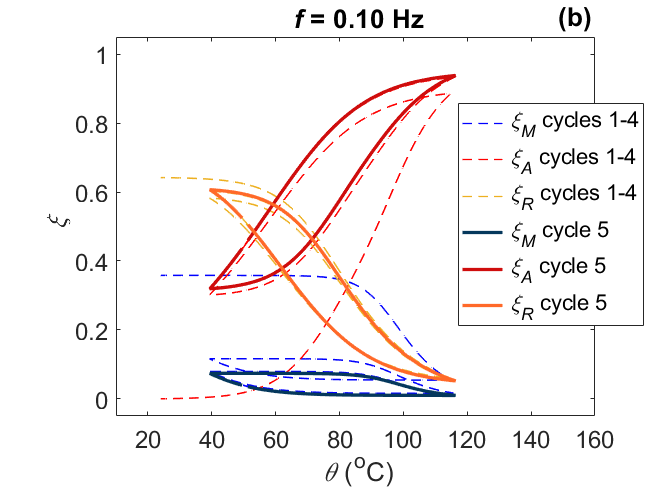} \\
        \includegraphics[scale=0.35]{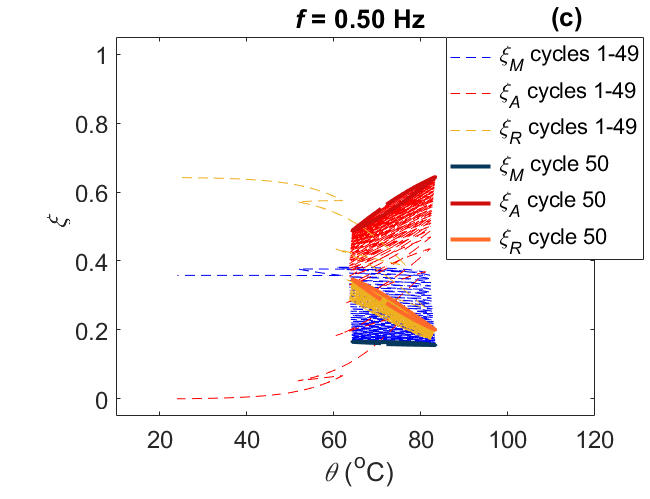} &
        \includegraphics[scale=0.35]{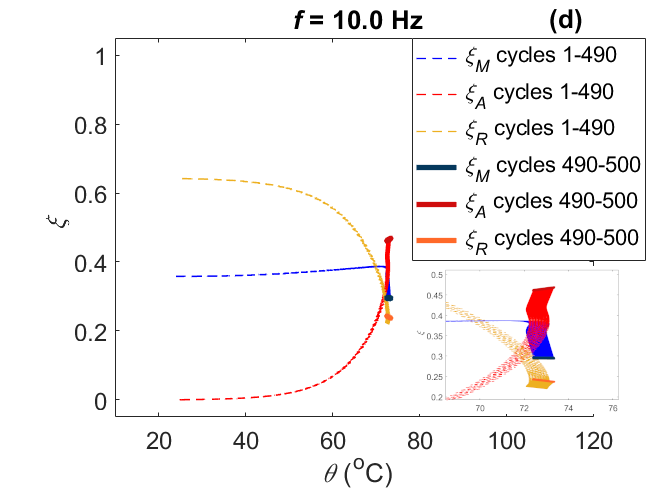}
    \end{tabular}
    \caption{Phase fractions \textit{vs.} sample temperature as predicted by the self-heating simulation runs. (a) $f=0.01$~Hz (b) $f=0.1$~Hz, (c) $f=1$~Hz, (d) $f=10$~Hz. The thin light-colored lines correspond to initial transient cycles and the thick dark-colored lines to the values close to or at the steady state.}
    \label{fig:KsivsT} 
\end{figure}

Fig.~\ref{fig:KsivsT} illustrates the evolution of the fractions $\xi_M,\xi_A$ and $\xi_R$ \textit{vs.} temperature $\theta$, as obtained by simulation. They correspond to the same current-driven runs and same number of cycles for each frequency as in figures~\ref{fig:rhovsIsimulation}b,d,f,h. The transient paths are depicted by thin, light-colored lines while the steady-state cycles use solid lines. This figure clearly demonstrates how the hysteresis in the $R$-$I$ and $V$-$I$ curves at small frequencies is explained by the hysteresis existing in the respective lattice phase proportion dynamics.

\begin{figure}
    \centering
    \begin{tabular}{c}
        \includegraphics[scale=0.50]{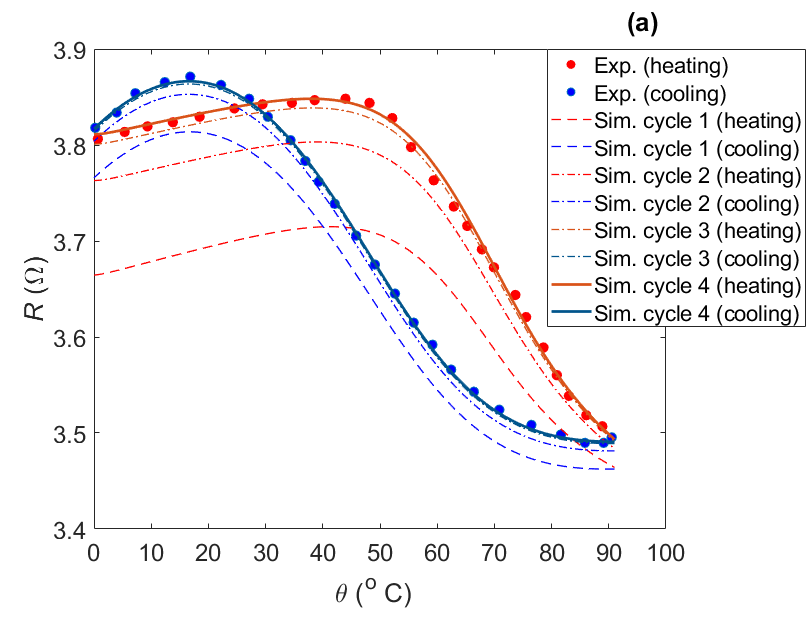} \\
        \includegraphics[scale=0.50]{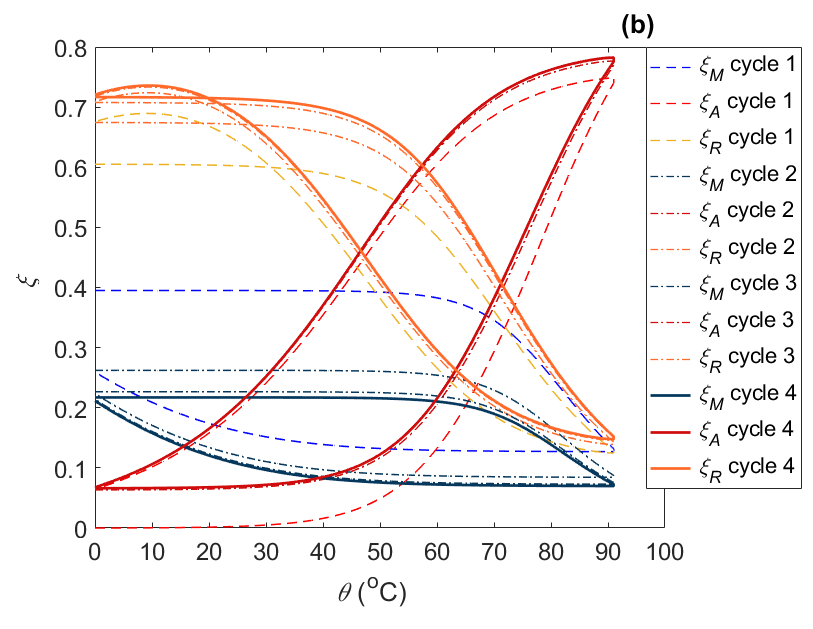}
    \end{tabular}
    \caption{Resistance $vs$. temperature for passive-heating experiments and simulation. Simulation results from four heating-cooling rounds are shown. (a) Resistance $vs$. temperature. (b) The respective variation of the three lattice phase fractions \textit{vs.} temperature for all four cycles as predicted by the model. The last cycle is shown with bold lines in both plots.}
    \label{fig:res_vs_temp} 
\end{figure}

Finally, comparison between numerical results and experimental data for the passive-heating experiments are shown in fig.~\ref{fig:res_vs_temp}a, where $R$ \textit{vs.} temperature $\theta$ ($^\circ$C) is plotted for four passive heating-cooling cycles. After three cycles the numerical results match the experimental data almost perfectly. The respective transition temperatures are shown in Table~\ref{tab:FreeParameterTableSTATIC}. Fig.~\ref{fig:KsivsT}b shows the respective cycles of the lattice phase fractions with respect to temperature. The hysteresis due to the M-to-A and A-to-R phase transitions, as well as the hysteresis due to R-to-M phase transition are evidently present and are the explanation of the two hysteretic curves seen at the resistance \textit{vs.} temperature data. 

\begin{table}[th]
    \centering
    \begin{tabular}{c|c||c|c||c|c}
    Parameter & Value & Parameter & Value & Parameter & Value\\ \hline
$\xi_M(t=0)$ &	0.358 &	$\xi_R(t=0)$ &	0.642 &	$\xi_A(t=0)$ &	0.000\\
$T_{s,MA}$ &	62.20 $^\circ$C &	$T_{f,MA}$ &	115.13 $^\circ$C	& $T_{c,MA}$ &	76.54 $^\circ$C \\
$T_{s,RA}$ &    44.93 $^\circ$C &   $T_{f,RA}$ &	114.95 $^\circ$C &   $T_{c,RA}$ &	63.90 $^\circ$C	\\
$T_{s,AR}$ &	121.24 $^\circ$C &	$T_{f,AR}$ &	30.78 $^\circ$C &	$T_{c,AR}$ &	55.29 $^\circ$C	\\
$T_{s,RM}$ & 113.68 $^\circ$C &   $T_{f,RM}$ &	-57.52 $^\circ$C &   $T_{c,RM}$ &	-11.135 $^\circ$C \\
$a_{MA}$ &	0.5  &	$a_{RA}$ &	0.5  \\
$a_{AR}$ &	0.5  &	$a_{RM}$ &	0.5
    \end{tabular}
    \caption{Fitted values of the model free parameters for the passive-heating simulations. Start and finish temperatures are calculated from \eqref{eq:TcMvsTsTf} for $c=0.05$.}
    \label{tab:FreeParameterTableSTATIC}
\end{table}

On the overall, we see that the model and the experimental data are very closely matched, as the model correctly reproduces the observed memristive hysteresis in both the $R$-$I$ and $V$-$I$ characteristics for all frequencies as well as its gradual disappearance with rising driving frequency.

\section{\label{sec:Discussion}Discussion and Concluding Remarks}

This paper describes models the thermally induced dynamical changes of NiTi resistivity based on the underlying dynamic changes of the lattice structural phase proportions (Martensite, Austenite and R-phase) that have been experimentally established for NiTi lattice structure in the past. The model was compared to previous electrical measurements performed with time-varying driving currents at various frequencies, as well as the experimental measurements of resistance at equilibrium temperature with passive heating. 

The comparisons have shown that the model reproduces accurately the shape of the hysteresis loop in both the $R$-$I$ and $V$-$I$ curves, for the current-driven, self-heated samples for both high and low frequencies. 
It was also demonstrated that the memristive electrical behavior of NiTi thin filaments can be adequately reproduced by a model that is based on the basic assumptions made for the present model, namely that: (i) the sample resistivity at each temperature is a linear combination of the resistivity of each of the three distinct lattice phases (Martensite, Austenite and R-phase), each phase having its own resistivity vs. temperature dependence; (ii) in general, the critical temperatures for the transitions between phases are different for each transition, possibly depending on whether the sample is at a heating or at a cooling state; and (c) thermal equilibrium is maintained during temperature changes.

Due to its dynamic character, the model can capture the full temporal evolution of the resistivities which is important when rapid changes in the lattice phase structure occur, especially at medium driving frequencies, where rich dynamics may be observed. The initial proportions of lattice phases play a significant role in the resistivity dynamics and the I-V characteristics, which is also a feature captured by the model.

Based on the experimental data currently available to us, which contained results from only one heating-cooling cycle for both the passive-heating experiments as well as the self-heating experiments for all driving current frequencies except for the $f=0.1$~Hz, for which approximately 2~1/2 cycles of data were available, we were able to determine the required model parameters with good precision, but there is still an ambiguity regarding the relative values of the critical temperatures which may be somewhat different when data from transient cycles and other frequency values are available. This is a subject of future work.

The present model does not include the possible dependence of resistivity on the atomic structure of interfaces in a NiTi polycrystal. As reported in a molecular dynamics simulation study of {NiTi} lattice phase transitions (\cite{KO201790}), during the transition process, there is a rapid build-up of inter-phase boundaries (interface between single crystal grains of different phases) and inter-granular boundaries (interface between grains of the same phase but with different spatial orientations) which cause an increase of the sample resistance. In the same study it is mentioned that inter-phase boundaries generally disappear after the end of a phase transition, when the sample is almost made up of a single lattice phase. However, inter-granular boundaries still exist among same-phase single crystals. In any case, temperature affects the relaxed atomic structure around boundaries. The change in relative position and structure of these boundaries as temperature changes and especially the migration of dislocations that are built around them may have an extra contribution to the variation of the resistance between heating rounds additional to the relative change of lattice phase proportions that are described by the present model. The possible contributions of the relaxed atomic structure around grain boundaries were inferred by the experimental study of Gori \textit{et al.} \cite{Gori06} (see Fig.~7 and the discussion therein), who suggested that resistivity at low temperatures depends on the (passive) heating-cooling cycle, even after cooling down to temperatures much lower than the temperature where the expected R-to-M phase transition is fully complete. However, the present study has shown that these resistivity relaxations, from cycle to cycle, can just as well be explained by the gradual change in the relative proportions between the R-phase, M-phase and A-phase, after each cycle. The existence of at least three phases in the alloy is a necessary condition to see this relaxation in the resistivity vs. temperature data, both in self-heating experiments where high-frequency temperature changes are present, as well as in the in passive-heating experiments where temperature varies slowly. Nevertheless, the possible dependence of resistivity on the atomic structure of inter-phase and inter-granular boundaries should be included in an upgraded future version of the model using, for instance, a statistical approach.

Regarding the small discrepancy found between the experimental data and numerical results for the $f=0.1$~Hz self-heating runs observed exclusively during the heating cycle, it could be either due to accidental experimental conditions not reported in \cite{stavrinides2019niti} (for instance the experiment not being conducted under the same external stress like the rest of the frequencies, a fact that would definitely lead to a lower start temperature of transition than the one fitting the other frequencies, or a different orientation of the wire relative to the horizontal leading to a different heat transfer coefficient value) or due to non-equilibrium effects due to the rapid heat input during the sharply rising currents at intermediate frequency values. The latter effects are not captured by the present model which assumes that free energy establishes a minimum at rates much greater than the temperature change rates involved. An important note should be made here: the width of the filament used, although small, is still at such a scale that the NiTi structure can be considered essentially as being 'bulk', \textit{i.e.} contributions to the free energy from surface effects and grain boundaries possibly suppressing the rate of phase transitions are still assumed to be negligible. Using filaments at nano-scale widths, for example, would probably make necessary that these effects are considered, a fact that may (among other things) impose the addition of non-equilibrium contributions to the model equations. The possibility that such non-equilibrium effects are the cause of the discrepancy between model and experimental data for $f=0.01$~Hz is still there. An extension of the model to include such effects is also a consideration for future work. Including stress/strain related electrical effects is another obvious extension to the model.

Finally, we note that a more detailed experimental study is needed for providing a more complete set of experimental data for better adjusting some of the parameters of the model. For example, one should combine DSC calorimetry results with resistivity measurements of the same samples conducted at precisely known environmental conditions in order to provide definite restrictions for the starting and finish temperatures of all the involved transitions.  

As a final conclusion, we would like to note that a multi-phase structure based model of NiTi resistivity that includes dynamical features, such as the one used here, when combined to electrical measurements under ac currents (or time-varying currents in general) over \textit{a wide range of frequencies}, can prove to be a very valuable tool in inferring various thermoelastic properties of a SMA, such as emmissivity (and its temperature dependence), resistivities of individual phases and their thermal dependence, latent heat production, specific heat capacity and its temperature dependence etc. A full demonstration of these capabilities by combining model with a multi-faceted experimental study would also be an interesting area for future work. 

%\begin{acknowledgments}
%We wish to acknowledge the support of the author community in usingREV\TeX{}, offering suggestions and encouragement, testing new versions, \dots.
%\end{acknowledgments}

%\nocite{*}

\bibliography{apssamp.bib}% Produces the bibliography via BibTeX.

\end{document}